\documentclass[preprint, trackchanges]{aastex701} 

\usepackage{multirow, url, color}
\usepackage{bbding, amssymb} 
\usepackage{ulem} 
\usepackage{verbatim} 
\usepackage{enumerate}
\usepackage{amsmath} 
\usepackage{booktabs} 
\usepackage{graphicx, subfigure}
\usepackage[fleqn]{mathtools}
\usepackage{mathrsfs}
\usepackage[makeroom]{cancel}

\newcommand{\textBoldadd}[1]{#1}
\newcommand{\textBoldmod}[2]{#2}
\newcommand{\textBoldrm}[1]{}

\usepackage{soul}
\soulregister\citep7
\soulregister\citet7
\soulregister\ref7
\soulregister\pageref7

\shorttitle{Modeling the 2023 February 24 SEP Event}
\shortauthors{Liu et al.}
\submitjournal{ApJ}

\begin{document}

\title{Counterintuitive Magnetic Connectivity and Energetic Particle Flux Differences among Nearby Spacecraft During the 2023 February 24 Solar Energetic Particle Event}

\correspondingauthor{Weihao Liu}
\email[show]{whliu@umich.edu}

\author[orcid=0000-0002-2873-5688]{Weihao Liu}
\affiliation{Department of Climate and Space Sciences and Engineering, University of Michigan, Ann Arbor, MI 48109, USA}
\email{whliu@umich.edu}
\author[orcid=0009-0009-2176-6017]{Xianyu Liu}
\affiliation{Department of Climate and Space Sciences and Engineering, University of Michigan, Ann Arbor, MI 48109, USA}
\email{xianyu@umich.edu}
\author[orcid=0000-0002-3176-8704]{David Lario}
\affiliation{Heliophysics Science Division, NASA Goddard Space Flight Center, Greenbelt, MD 20771, USA}
\email{david.larioloyo@nasa.gov}
\author[orcid=0000-0003-3936-5288]{Lulu Zhao}
\affiliation{Department of Climate and Space Sciences and Engineering, University of Michigan, Ann Arbor, MI 48109, USA}
\email{zhlulu@umich.edu}
\author[orcid=0000-0001-9360-4951]{Tamas I. Gombosi}
\affiliation{Department of Climate and Space Sciences and Engineering, University of Michigan, Ann Arbor, MI 48109, USA}
\email{tamas@umich.edu}
\author[orcid=0000-0001-7420-6405]{Alexander D. Shane}
\affiliation{Department of Climate and Space Sciences and Engineering, University of Michigan, Ann Arbor, MI 48109, USA}
\affiliation{Department of Aerospace Engineering Sciences, University of Colorado, Boulder, CO 80309, USA}
\email{adshane@umich.edu}
\author[orcid=0000-0002-6118-0469]{Igor V. Sokolov}
\affiliation{Department of Climate and Space Sciences and Engineering, University of Michigan, Ann Arbor, MI 48109, USA}
\email{igorsok@umich.edu}

\begin{abstract}
For solar energetic particles (SEPs), it is generally expected that observers magnetically closer to the eruption source region exhibit higher particle intensities than those poorly connected to the eruption site. However, the 2023 February 24 SEP event departs from this simple picture: Earth and STA, near 1 au, are nominally better connected to the source region, whereas Solar Orbiter (SolO), at 0.77 au but less favorably connected, observed SEP fluxes more than an order of magnitude higher. This difference cannot be simply explained by nominal magnetic connectivity or radial scaling of SEP fluxes alone. 
To investigate this behavior, we perform a global magnetohydrodynamic simulation of the associated coronal mass ejection (CME) using the Alfv\'{e}n Wave Solar-atmosphere Model-Realtime (AWSoM-R). The simulation reveals that the CME flux rope originates close to a coronal streamer and as it propagates and expands, the CME-driven shock is effectively distorted, developing into two distinct flanks with different strengths. 
Although the three spacecraft are separated by only $\lesssim$32$^{\circ}$ in heliolongitude, their magnetic footpoints differ by $\gtrsim$50$^{\circ}$ in longitude because of a nearby stream interaction region. Specifically, Earth and STA connect to a weaker shock region, while SolO connects to the shock nose with a higher compression ratio and more efficient particle acceleration. 
We further simulate SEPs using the Multiple-Field-Line Advection Model for Particle Acceleration (M-FLAMPA) coupled with AWSoM-R, obtaining results that reproduce the observed flux differences among the three spacecraft, demonstrating that this counterintuitive behavior results from their connections to different regions of the inhomogeneous CME-driven shock. 
\end{abstract}


\section{Introduction} \label{sec1:intro}

    \subsection{Background} \label{sec1.1:bg}

    Solar energetic particles (SEPs) are high-energy ions and electrons accelerated during solar eruptive events and transported through the heliosphere along the interplanetary magnetic field (IMF). Among all SEP events, large gradual SEP events are particularly hazardous because they can produce intense particle fluxes with extended decay phases, often lasting several days, thereby posing significant radiation risks to spacecraft systems and astronauts \citep[e.g.,][]{desai2016large, papaioannou2025predicting}. 
    These events are commonly associated with fast coronal mass ejections (CMEs), where CME-driven shocks provide spatially extended acceleration sources as they propagate from the low corona into interplanetary space \citep{reames1999particle, klein2017acc}. Therefore, the SEP intensity at a given location depends not only on the geometry and strength of the shock \citep{giacalone2005particle, tylka2005shock, guo2015acceleration, chen2022solar}, but also on how this location is magnetically connected to the shock region where particles are accelerated \citep{cane1988role, lario1998energetic, kouloumvakos2019connecting, wijsen2022observation, liu2025physics, chen2025evidence, ding2025investigation}. Magnetic connectivity is thus central to interpreting SEP onset times, peak intensities, spectral properties, and the longitudinal distribution of particles in multi-spacecraft observations \citep[e.g.,][]{richardson201425, lario2013longitudinal, lario2017link, xie2019statistical, kennis2024magnetic, zhou2026three}. 

    A common expectation is that spacecraft more closely magnetically connected to the eruption source region should observe earlier onsets and more intense particle fluxes than those poorly connected \citep[e.g.,][]{van1975variation, lario2006radial, lario2013longitudinal, richardson201425}. In a nominal Parker-spiral approximation, this connectivity is estimated from the observer longitude, solar wind speed, and source region longitude \citep{parker1958dynamics}. 
    \textBoldadd{To relate the Parker-spiral connectivity in the heliosphere to the low-coronal source region, the magnetic field line is often traced ballistically to a source surface and then mapped to the photosphere using a Potential Field Source Surface (PFSS) model \citep[e.g.,][]{altschuler1969magnetic, schatten1969model, gieseler2023solar, biro2025developing}.}
    Although actual SEP observations are affected by the evolving shock geometry and particle transport effects such as cross-field diffusion, magnetic connectivity remains one of the most important factors used to interpret observer-to-observer differences in SEP fluxes \citep[see SEP event studies by][]{rouillard2012longitudinal, zhu2021shock, khoo2024multispacecraft, dresing2025reason, tao2025simulation, lario2026major}. 

    \subsection{The 2023 February 24 SEP Event: Overview} \label{sec1.2:overview}

\begin{figure*}[tp!]
    \centering{
    \includegraphics[width=0.975\textwidth]{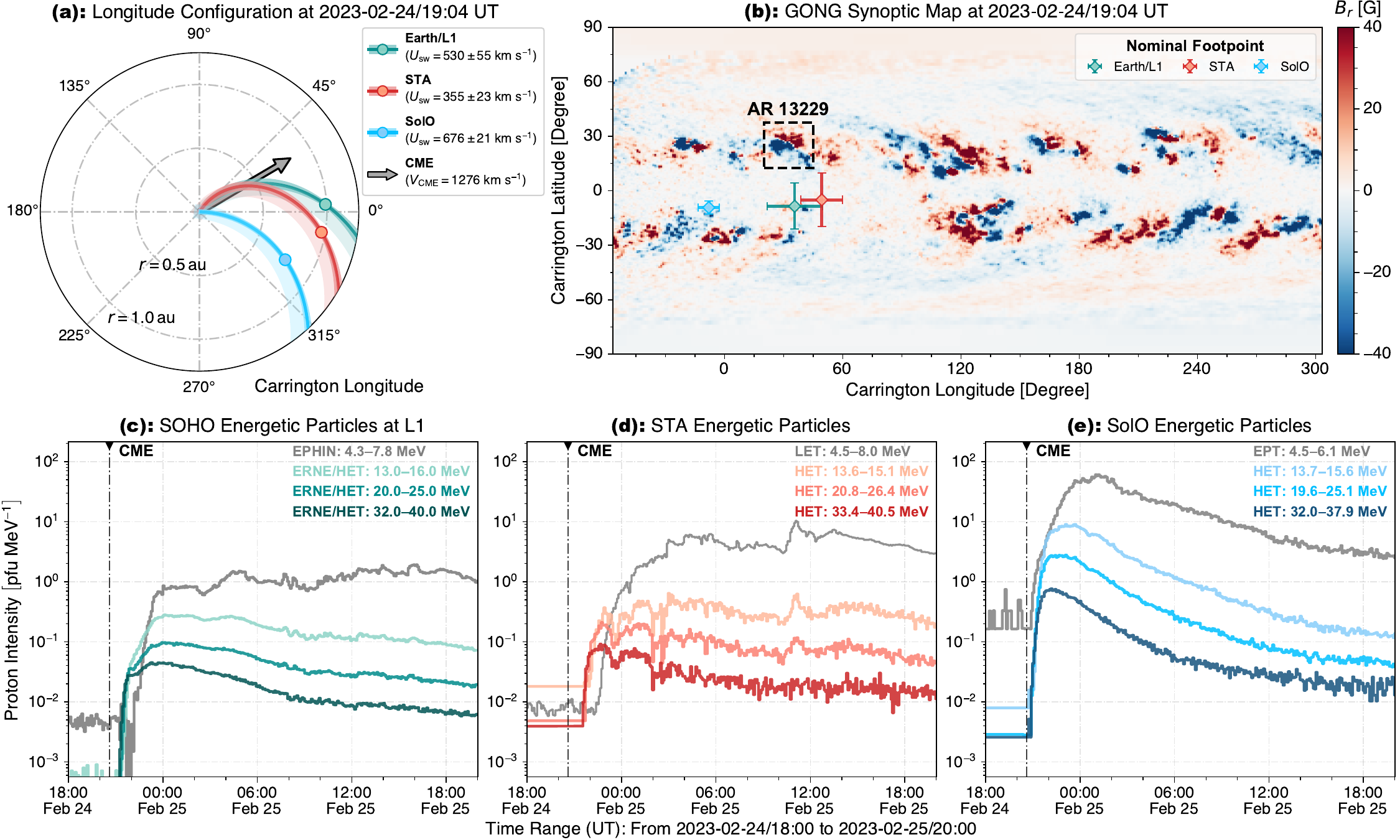}}
    \caption{Observational overview of the 2023 February 24 SEP event. 
    (a) Carrington-longitude configuration of Earth/L1 (green circle), STA (red circle), and SolO (blue circle). Curves in the corresponding colors show the nominal IMF lines estimated \textBoldmod{from}{using the 1-day pre-eruption mean} local solar wind speeds, \textBoldmod{and the}{with the light-colored bands indicating the uncertainty associated with varying each speed by its corresponding standard deviation and the following 1-day corotation-related location difference; the legend lists the mean and standard deviation local solar wind speed values. The} black arrow indicates the CME propagation direction. 
    (b) GONG synoptic map of the photospheric radial magnetic field ($B_r$) at 19:04 UT on 2023 February 24, shortly before the eruption, with the eruption source region, AR 13229, marked by the dashed black box. \textBoldadd{The diamond symbols indicate the nominal footpoints of Earth/L1, STA, and SolO estimated from the local solar wind speeds shown in panel (a) and the PFSS model; the horizontal and vertical bars indicate the corresponding footpoint uncertainties associated with the standard deviation of the solar wind speed estimate and the 1-day corotation-related location differences.} 
    (c)–(e) Proton time--intensity profiles observed by SOHO/EPHIN and ERNE at L1, STA/LET and HET, and SolO/EPD/EPT and HET, respectively. Here, ``pfu" is an acronym of particle flux unit, defined as $1\;\mathrm{pfu} = 1\;\mathrm{count \; cm^{-2}\; s^{-1}\; sr^{-1}}$. Four representative energy channels from $\sim$4 to $\sim$40 MeV are shown for each observer, selected to approximately match each energy range among different spacecraft. The vertical dashed-dotted black lines mark the CME eruption time.} \label{fig1:obs}
    \end{figure*}

    An SEP event on 2023 February 24 provides a counterintuitive example that challenges the simple connectivity-based expectation described in Section \ref{sec1.1:bg}. 
    Figure~\ref{fig1:obs} summarizes the SEP observations and the longitudinal distribution of spacecraft during this SEP event, together with the site of the source region, NOAA active region (AR) 13229, marked on the synoptic map from the Global Oscillation Network Group \citep[GONG\footnote{\url{https://nso.edu/data/nisp-data/magnetograms/} \label{ftn:gong}},][]{harvey1996global, hill2018global}. 
    
    On 2023 February 24, AR 13229 at Stonyhurst heliographic coordinates N29W24 launched an M3.7 X-ray flare followed by a large filament eruption\footnote{Details are available at \url{http://svs.gsfc.nasa.gov/5082}.}. The soft X-ray emission began at 20:03 UT, peaked at 20:30 UT, and ended at 21:29 UT. The associated CME was first observed by the Large Angle and Spectrometric Coronagraph \citep[LASCO\footnote{\url{https://cdaw.gsfc.nasa.gov/CME_list} \label{ftn:lascocme}},][]{brueckner1995lasco, yashiro2004catalog} on board the SOlar and Heliospheric Observatory \citep[SOHO,][]{domingo1995soho} at 20:36 UT, with the leading edge at a position angle\footnote{The position angle is measured in the plane-of-sky coronagraph images counter-clockwise from the solar north.} (PA) of $\sim$344$^{\circ}$ at a height of 3.60 solar radii ($R_\mathrm{s}$). 
    The CME was also observed by coronagraphs, COR1 and COR2, of the Sun Earth Connection Coronal and Heliospheric Investigation \citep[SECCHI\footnote{See \url{https://cor1.gsfc.nasa.gov/catalog/cme/2023/HongXie_COR1_preliminary_event_list_2023-02.html}. \label{ftn:coracme}},][]{howard2008sun} on board the Solar Terrestrial Relations Observatory-Ahead \citep[STA,][]{kaiser2008stereo}. It first appeared in COR1 at 20:36 UT and was marginally visible in COR2 as early as 20:38 UT at PA $\simeq 322^{\circ}$, and became clearly visible in COR2 at 20:53 UT. 
    Combining multi-spacecraft coronagraph observations, the CME leading-edge speed is estimated to be $1276 \;\mathrm{km \;s^{-1}}$ after correcting for projection effects, but this estimate does not include the surrounding faint shock halo, as reported in the Space Weather Database Of Notifications, Knowledge, Information (DONKI) database\footnote{\url{https://kauai.ccmc.gsfc.nasa.gov/DONKI/search/} \label{ftn:donki}}. 
    
    This SEP event is well observed by SOHO at the Sun-Earth Lagrangian point L1, by STA 13$^{\circ}$ east of Earth, and Solar Orbiter \citep[SolO,][]{muller2020solar} 32$^{\circ}$ east of Earth. As shown in Figure~\ref{fig1:obs}(a), their locations in Carrington heliographic coordinates, given as the heliocentric distance, longitude, and latitude, are (0.99 au, $2.5^{\circ}$, $-7.1^{\circ}$) for Earth, (0.97 au, $349.7^{\circ}$, $-6.8^{\circ}$) for STA, and (0.77 au, $330.0^{\circ}$, $-2.2^{\circ}$) for SolO. The three observers are relatively close in heliolongitude, with a maximal longitudinal separation of $\sim$32$^\circ$, but their nominal \textBoldrm{Parker-spiral} footpoints differ by $\gtrsim$50$^{\circ}$ \textBoldadd{primarily} because of the different local solar wind speeds, which are estimated using a 1-day window before the CME eruption and listed in the legend of Figure~\ref{fig1:obs}(a). 
    \textBoldadd{For this nominal estimate, the field-line path is estimated ballistically from each spacecraft to $2.5\;R_\mathrm{s}$ using the corresponding local solar wind speed, and then mapped from $2.5\;R_\mathrm{s}$ to the photosphere using the PFSS model \citep[e.g.,][]{gieseler2023solar, biro2025developing}, yielding the nominal footpoints shown in Figure~\ref{fig1:obs}(b).} 
    Specifically, \textBoldadd{the three spacecraft's footpoints are close in latitude, while in longitude,} Earth and STA \textBoldmod{are nominally}{appear} better connected to the source region than SolO; however, the SEP observations show that SolO detects much higher particle intensities. 
    
    Figures~\ref{fig1:obs}(c)–(e) show the time--intensity SEP profiles measured by (c) the Energetic and Relativistic Nuclei and Electron instrument \citep[ERNE,][]{torsti1995energetic, valtonen1997energetic} and the Electron Proton Helium Instrument \citep[EPHIN,][]{muller1995costep} on board SOHO at L1; (d) the Low Energy Telescope \citep[LET,][]{mewaldt2008low} and High Energy Telescope \citep[HET,][]{von2008high} on board STA; and (e) the Electron Proton Telescope (EPT) and HET of the Energetic Particle Detector \citep[EPD,][]{rodriguez2020energetic} suite of instruments on board SolO. 
    We note that the DONKI database reports four additional small SEP-associated eruptions within the following day, before the onset of a second large SEP event occurring at $\sim$19:36 UT on 2023 February 25. Hence, our following analysis focuses on the February 24 CME eruption and its associated SEP evolution, before the later event may significantly affect the particle intensities. 
    As displayed in Figures~\ref{fig1:obs}(c)–(e), SolO observes SEP fluxes more than an order of magnitude higher than those measured by SOHO and STA across four representative energy channels ranging from $\sim$4 MeV to $\sim$40 MeV. Note that if the particle flux scales with heliocentric radial distance $r$ as $r^{-3}$ or $r^{-2}$ \citep[e.g.,][]{lario2006radial, cao2025radial}, the expected flux ratio between 0.77 au and 1.0 au is only $\sim$2.2 or $\sim$1.7, respectively. 
    Therefore, this unexpected behavior cannot be simply explained by nominal \textBoldrm{Parker-spiral} connectivity or usual radial scaling of SEP fluxes. 


    \subsection{Motivation and Paper Outline} \label{sec1.3:outline}

    Motivated by the counterintuitive discrepancy described in Section~\ref{sec1.2:overview}, the key question of this study is why SolO observed substantially higher SEP fluxes than SOHO and STA despite the modest heliolongitudinal and radial separations among the three observers. 
    To investigate the root cause of this difference, we (i) perform global magnetohydrodynamic (MHD) simulations of the background solar wind and CME propagation, (ii) apply a shock-capturing tool to analyze the shock properties and magnetic connectivity of each observer, and (iii) simulate the SEP distribution function, using the SOlar wind with FIeld lines and Energetic particles \citep[SOFIE\footnote{Available at \url{https://github.com/SWMFsoftware} and \url{https://ccmc.gsfc.nasa.gov/ror/requests/SH/SWMF-AWSoM/swmfawsom_user_registration.php}. \label{ftn:sofie}},][]{zhao2024solar, liu2025physics, liu2026testbed} model, an integrated solar corona–interplanetary medium–CME–SEP model suite. In the following, we describe each modeling component of SOFIE and present the corresponding results: Section~\ref{sec2:mhd} presents the MHD simulation and shock-connectivity analysis, Section~\ref{sec3:sep} describes the SEP simulation results, and Section~\ref{sec5:sumcon} summarizes the main findings of this study.

\section{MHD Simulations} \label{sec2:mhd}

    \subsection{The AWSoM-R and Background Solar Wind} \label{sec2.1:sw}
    
    To model the global coronal and heliospheric environment, we first construct a steady-state background solar wind using the Alfv\'{e}n Wave Solar atmosphere Model-Realtime \cite[AWSoM-R,][]{sokolov2013magnetohydrodynamic, sokolov2021threaded, van2014alfven}. AWSoM-R involves Alfv\'{e}n-wave-driven coronal heating and solar wind acceleration, where the coronal heating is due to dissipation of two counter-propagating Alfv\'{e}n-wave turbulence populations along the magnetic field. The model includes physically consistent treatments of wave reflection, turbulent dissipation, electron--proton heat partitioning, electron heat conduction, and radiative cooling, providing a self-consistent description of the coronal and heliospheric plasma and magnetic-field structure. Our model also employs a stream-aligned MHD treatment, as described in \cite{sokolov2022stream}, to better preserve magnetic connectivity and reduce artificial numerical reconnection that can lead to unphysical ``V-shaped'' magnetic field lines. 
    
    We use the hourly updated, zero-point corrected GONG synoptic map shortly before the CME eruption (see footnote \ref{ftn:gong} for the data source), as shown in Figure~\ref{fig1:obs}(b), to drive a steady-state simulation of the corona. We then use this solution as the background state into which the CME is launched (Section~\ref{sec2.2:cme}) and through which SEPs propagate (Section~\ref{sec3:sep}). In this event, we adopt a Poynting flux parameter, a major free parameter of AWSoM-R, of $0.4~\mathrm{MW\;m^{-2}\;T^{-1}}$ based on the studies of \cite{jivani2023global} and \cite{huang2024solar}. 
    
    The simulation domain consists of a threaded-field-line region from 1.0 to 1.1 $R_\mathrm{s}$, a block-adaptive three-dimensional (3D) spherical grid in the solar corona (SC) from 1.1 to 24 $R_\mathrm{s}$, and a block-adaptive Cartesian grid in the inner heliosphere (IH) from 20 to $650 \;R_\mathrm{s}$. In the region between 1.0 and 1.1 $R_\mathrm{s}$, AWSoM-R employs the threaded-field-line model proposed by \cite{sokolov2021threaded} and solves the field-aligned hydrodynamic equations\textBoldadd{, with the magnetic field extracted from the PFSS model expanded to spherical-harmonic degree 180  \citep{schatten1969model, altschuler1969magnetic, toth2011obtaining}}. In the SC and IH domains, AWSoM-R solves the global extended MHD equations as detailed in \cite{gombosi2018extended}. The SC and IH domains overlap between 20 and $24 \;R_\mathrm{s}$, where a buffer region couples the SC and IH solutions. 
    In the SC domain, the initial angular resolution is set to 1.4 degrees in both longitude and latitude. The resolution is then refined by one level to 0.7 degrees within $r = 1.7 \;R_\mathrm{s}$, and coarsened by one level to 2.8 degrees beyond this heliocentric distance. 
    We apply adaptive mesh refinement \citep[AMR,][]{gombosi2003adaptive, toth2012adaptive} to resolve the heliospheric current sheet (HCS), cone regions directed toward Earth, STA, and SolO, and later along the CME propagation direction in both SC and IH domains, ensuring that key coronal and heliospheric structures are well captured. The refined grid contains $\sim$4 million and $\sim$150 million cells in the SC and IH domains, respectively.
    
    \begin{figure*}[tp!]
    \centering{
    \includegraphics[width=0.95\textwidth]{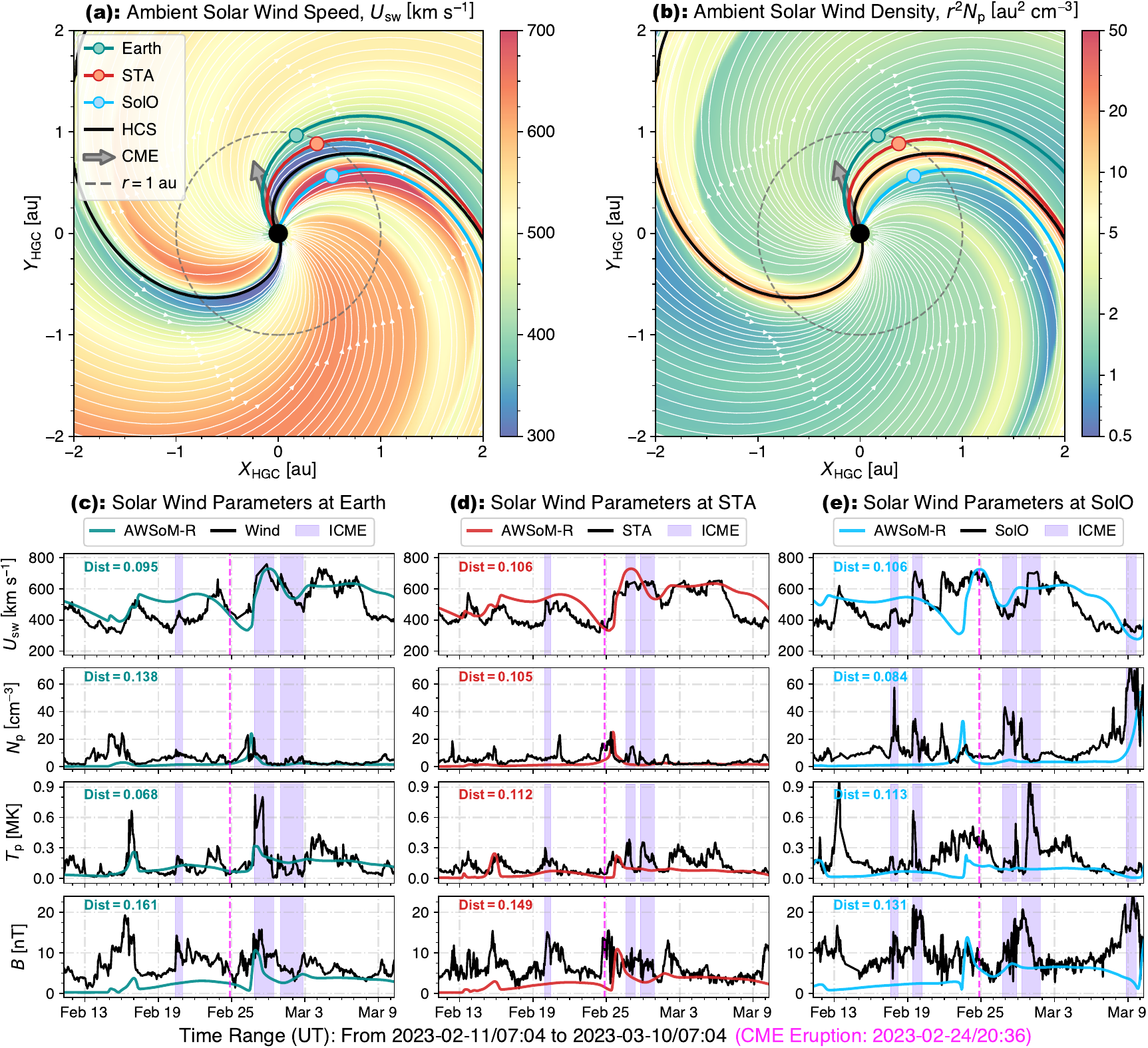}}
    \caption{Simulated background solar wind. 
    (a) Ambient solar wind speed ($U_\mathrm{sw}$) and (b) density scaled by the heliocentric distance ($r^2N_\mathrm{p}$) in the solar equatorial plane with Carrington heliographic (HGC) coordinates used. White curves with arrows represent magnetic field lines, colored curves show the field lines connected to Earth (green), STA (red), and SolO (blue), and the black curve marks the HCS. \textBoldadd{The gray arrow indicates the initial CME propagation direction, as also shown in Figure \ref{fig1:obs}(a) in a different coordinate system.} The gray dashed circle indicates $r = 1$ au, and the black solid circle at the center is the inner boundary of the IH domain at $r = 20 \;R_\mathrm{s}$. 
    (c)–(e) Comparison between the AWSoM-R results (colored curves) and \textit{in-situ} solar wind observations (black curves) at Earth, STA, and SolO, respectively, over a 27-day period centered at 19:04 UT on 2023 February 24. From top to bottom, the panels show solar wind speed, proton density, proton temperature, and magnetic field strength, each with an annotated distance value (``Dist") quantifying the model--observation difference. The vertical magenta dashed lines mark the CME eruption time of this event, and the purple shaded regions indicate all identified ICME intervals within this 27-day window.} \label{fig2:sw}
    \end{figure*}
    
    Figure~\ref{fig2:sw} shows the modeled ambient solar wind and heliospheric environment shortly before the onset of this SEP event, at 19:04 UT on 2023 February 24. In the vicinity of Earth, STA, and SolO, the solar wind background is highly structured by the HCS (indicated by the thick black line in Figures~\ref{fig2:sw}(a) and \ref{fig2:sw}(b)) and a nearby stream interaction region (SIR), where fast solar wind interacts with slower upstream wind and forms a compressed plasma and magnetic-field structure located in the yellow regions of panel (b) followed by high-speed solar wind streams (red regions in panel (a)). \textBoldadd{The gray arrows in panels~(a) and (b) identify the initial CME propagation direction, which lies close to the HCS and the associated SIR structure; the resulting early CME evolution and CME–SIR interaction are examined later in Section~\ref{sec2.2:cme}. Here,} 
    SolO is located in a relatively low-density region and high-speed solar wind stream, just after the passage of the HCS; in fact, a sector-boundary crossing was observed \textit{in situ} around 18:00 UT on 2023 February 21, although not shown here. Earth transitioned from fast to slow solar wind throughout February 24 (panel (c)), whereas STA observed the SIR arriving near the end of February 24 (panel (d)). As a result, this HCS–SIR structure strongly affects the observer connectivity, where the MHD field lines connected to Earth, STA, and SolO are deflected and separated by the SIR, leading to substantially different magnetic footpoints and later connections to different portions of the CME-driven shock. 
    
    Figures~\ref{fig2:sw}(c)–(e) validate the modeled background solar wind by comparing AWSoM-R results with \textit{in-situ} solar wind observations by the Wind spacecraft \citep{harten1995design} at L1 near Earth \textBoldadd{\citep[obtained from][]{cdaw_OMNIsw}}, STA \textBoldadd{\citep[obtained from][]{cdaw_STAsw}}, and SolO\footnote{\textBoldadd{Obtained from \url{https://cdaweb.gsfc.nasa.gov/pub/data/solar-orbiter/coho1hr_magplasma/}.}} over a 27-day period centered on this event. The vertical magenta dashed line in each panel marks the CME eruption time for reference. Overall, AWSoM-R captures the large-scale variations in solar wind parameters and magnetic field strength at all three locations, especially the SIR around the eruption time, although discrepancies remain in the precise arrival time of heliospheric structures and during interplanetary CME (ICME) intervals, which are not expected to be fully reproduced by a steady-state global background solar wind simulation \cite[e.g.,][]{sachdeva2019validation, sachdeva2023solar, zhao2024solar, koban2025dailyAGU}. Note that the ICME intervals plotted in Figures~\ref{fig2:sw}(c)–(e) are first obtained from publicly available catalogs for Wind\footnote{\url{https://wind.nasa.gov/ICME_catalog/ICME_catalog_viewer.php} \label{ftn:windcat}} \citep{nieves2018understanding}, STA\footnote{\url{https://stereo-ssc.nascom.nasa.gov/data/ins_data/impact/level3/} \label{ftn:stacat}} \citep{jian2018stereo}, and SolO\footnote{\url{https://science.gsfc.nasa.gov/lassos/ICME_catalogs/solo-catalog.shtml} \label{ftn:solocat}} \citep{nieves2016circular}, together with a comprehensive ICME catalog, ICMECAT\footnote{See \url{https://helioforecast.space/icmecat}\textBoldrm{; version 24, used in this study, is available at \url{https://figshare.com/articles/dataset/HELCATS_Interplanetary_Coronal_Mass_Ejection_Catalog_v2_0/6356420}}. \label{ftn:icmecat}} \textBoldadd{\citep{mostl2025icmecat, mostl2026magnetic},} and are then visually inspected and confirmed. 
    We further quantify the model--observation agreement outside the ICME intervals using the data-distance metric introduced in \citet{sachdeva2019validation}, which measures the difference between the simulated and observed \textit{in situ} time profiles and is indicated by the ``Dist" values in Figures~\ref{fig2:sw}(c)–(e). Previous validation studies have shown that well-reproduced solar wind profiles typically yield distance values of $\lesssim 0.15$, whereas clear deviations over a 27-day window correspond to values of $\gtrsim 0.2$ \citep[e.g.,][]{sachdeva2019validation, huang2024solar, koban2025validation}. 
    In our case, the distance values are relatively small across all four parameters and three observers, indicating that the modeled solar wind provides a reasonable background for analyzing the magnetic connectivity and the subsequent CME and SEP propagation in this event. 
    
    \begin{figure*}[tp!]
    \centering{
    \includegraphics[width=0.7\hsize]{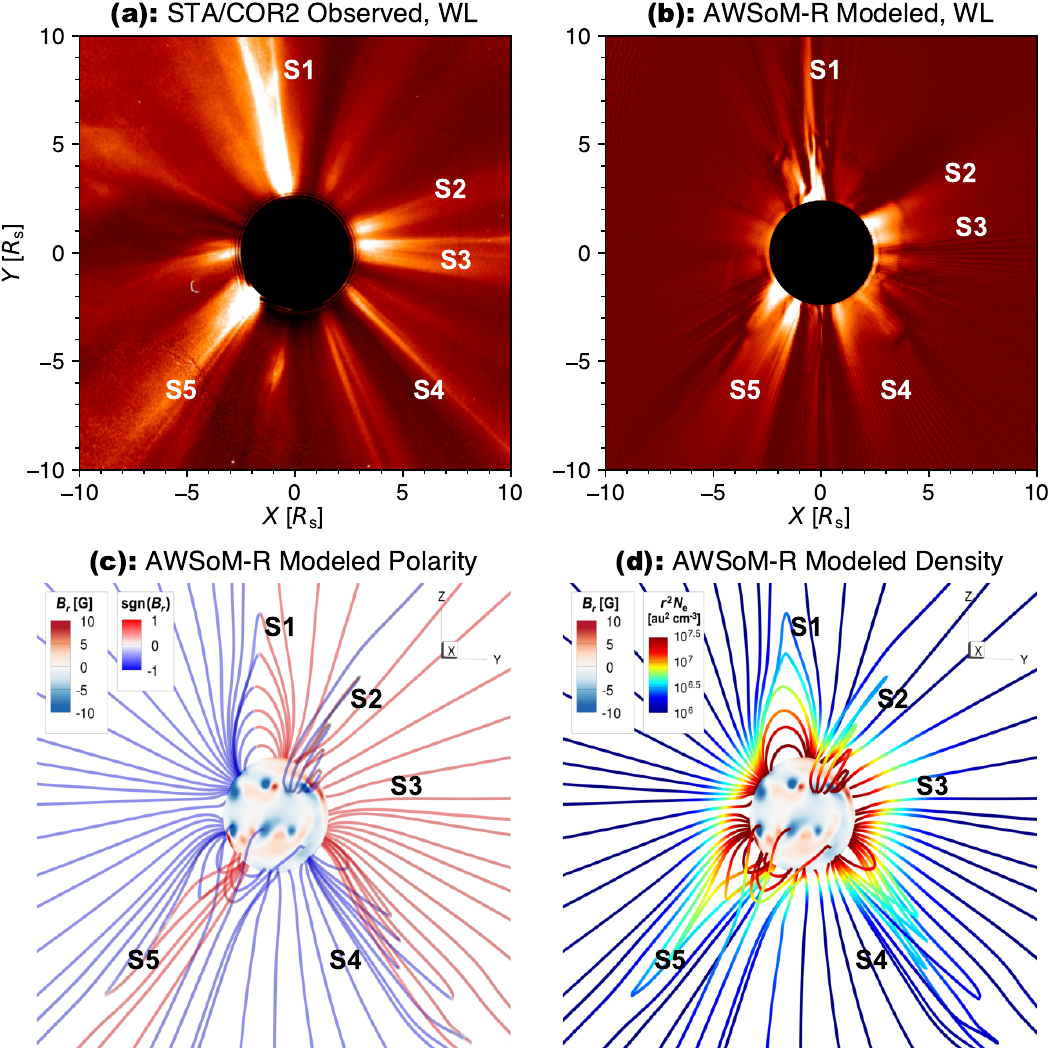}}
    \caption{Steady-state coronal structure from the STA viewpoint.
    (a) Pre-event STA/COR2 WL observation, at 20:23 UT on 2023 February 24, with five main streamer structures labeled S1–S5 clockwise from the north. 
    (b) Corresponding WL image synthesized from the AWSoM-R modeled background solar wind. The central black disk in panels (a) and (b) marks the inner boundary of the COR2 field of view at $r = 2.5 \;R_\mathrm{s}$. 
    (c) The AWSoM-R modeled magnetic polarity in 3D space. The SC inner boundary at $r = 1.1 \;R_\mathrm{s}$ is colored by the radial magnetic field strength ($B_r$), and selected magnetic field lines are colored by their polarity, $\mathrm{sgn}(B_r)$. 
    (d) Same as panel (c) except that the magnetic field lines are colored by the electron density scaled by heliocentric distance ($r^2N_\mathrm{e}$) on a logarithmic scale. 
    S1--S5 are also labeled in panels (b)–(d) to identify the corresponding modeled structures used for comparison.} \label{fig3:bgwl}
    \end{figure*}
    
    Moreover, we validate the modeled coronal structures by synthesizing a white-light (WL) image from the steady-state AWSoM-R simulation and comparing it with the pre-event coronagraph observations. The synthetic WL image is processed using an unsharp-mask filter and an exponential radial filter as described in Section 3.1 of \cite{liu2026simulating}. Note that the filters for the synthetic WL image are different from those used for the STA/COR2 observations, which are obtained directly from the Integrated Space Weather Analysis \citep[ISWA,][]{maddox2010utilizing} system\footnote{\url{https://iswa.ccmc.gsfc.nasa.gov/iswa_data_tree/observation/solar/soho/} \label{ftn:iswa}}. 
    As shown in Figures~\ref{fig3:bgwl}(a) and \ref{fig3:bgwl}(b), the synthetic WL image (right) reproduces the large-scale structures observed by STA/COR2 (left), labeled S1–S5, although the observed structures appear generally sharper and more filamentary. 
    The modeled 3D magnetic polarity and scaled electron density from the STA viewpoint, shown in Figures~\ref{fig3:bgwl}(c) and \ref{fig3:bgwl}(d), further indicate that these structures are associated with the large-scale magnetic polarity and enhanced coronal density. Besides, Figure~\ref{fig3:bgwl}(c) suggests that the HCS is approximately north-south oriented during this event\textBoldadd{, as indicated by the north-to-south elongated transition between the opposite magnetic polarity regions near the Sun}. 
    Among the five labeled structures, S1, S2, S4, and S5 are identified as helmet streamers, while S3 is identified as an open-flux region near and beyond the west limb. In particular, the magnetic field lines of the streamer S2 are anchored near the eruption AR discussed later in Section~\ref{sec2.2:cme}, which strongly suggests that the subsequent CME--streamer interaction involves the streamer S2. 
    
    A modest discrepancy appears near the east limb in the synthetic WL image, where the modeled streamer structure differs from the STA/COR2 observation, appearing to correspond to an open-flux region (see panels (c) and (d)). This is mainly because this portion of the synoptic map is constrained by older magnetogram data; nevertheless, this region is far from the eruption site and does not play a significant role in the CME propagation analyzed in this work. 
    Recent studies have also investigated the sensitivity of steady-state streamer structures and/or CME eruptions to different synoptic maps and far-side magnetic field constraints \citep[e.g.,][]{ledvina2023modeling, perri2024impact, shi2025role, liu2026simulating, baratashvili2026modelling, sachdeva2026evolution}. Overall, despite some small-scale and far-limb discrepancies, the modeled background captures the dominant coronal density distribution and magnetic topology relevant to the subsequent CME propagation.

    \subsection{CME Flux Rope Generation and Propagation} \label{sec2.2:cme}
    
    \begin{table}[tp!]
    \begin{center}
    \caption{GL flux rope parameters used for the CME simulation.} \label{tab1:param}
    \setlength\tabcolsep{15pt}{
    \begin{tabular}{lc}
    \hline\hline 
        Parameter & Value \\ 
    \hline 
        Type & Spheromak \\
        Source region location\tablenotemark{$*$} & $\left(32.5^{\circ},\, 25.4^{\circ}\right)$ \\
        Major axis orientation & $268.51^{\circ}$ \\
        Radius & $0.59 \;R_\mathrm{s}$ \\
        Stretch & $0.60 \;R_\mathrm{s}$ \\
        Apex height & $0.79 \;R_\mathrm{s}$ \\
        Magnetic field strength & $16.93 \;\mathrm{G}$ \\
        CME speed & $1276 \;\mathrm{km\; s^{-1}}$ \\
    \hline
    \end{tabular}}
    \end{center}
    \tablenotetext{*}{ This location is given as the Carrington longitude and latitude. }
    \end{table}
    
\begin{figure*}[tp!]
    \centering{
    \includegraphics[width=0.975\hsize]{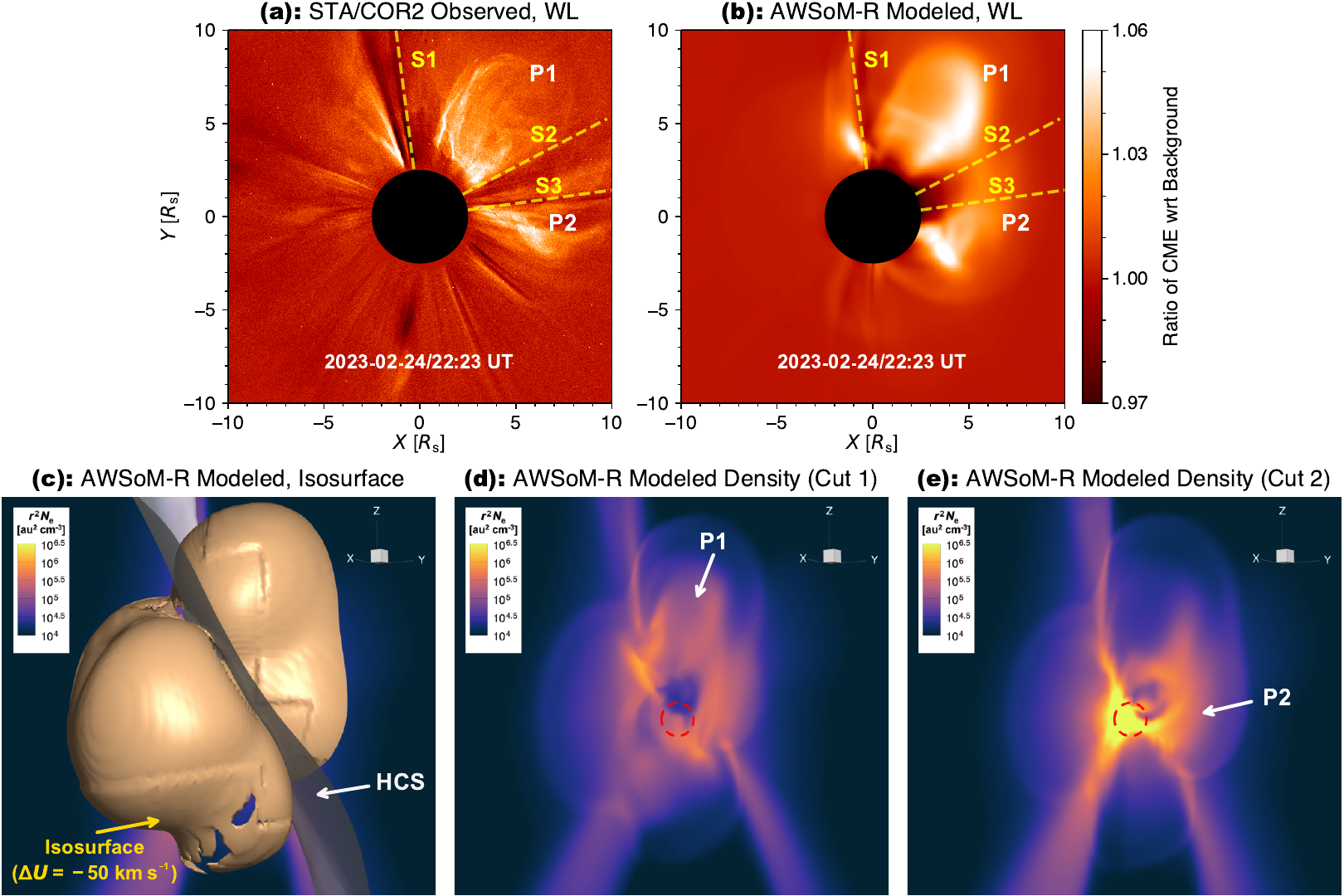}}
    \caption{CME--streamer interaction during the early CME propagation.
    (a)(b) WL images observed by STA/COR2 and synthesized from the AWSoM-R simulation, respectively, at 22:23 UT on 2023 February 24. These images are colored by the change in WL total brightness, shown as the ratio of the CME image to the pre-event background. The black disk in the center marks the inner boundary of the COR2 field of view at $r = 2.5 \;R_\mathrm{s}$. Labels P1 and P2 denote the two main propagation fronts of the deformed CME structure. \textBoldadd{The yellow dashed lines mark the locations of S1–S3 identified in Figure~\ref{fig3:bgwl}, following the position angles of their maximum WL intensity in the pre-eruption corona.} 
    (c) 3D view of the AWSoM-R modeled CME disturbance at the same time as panels (a) and (b), shown by the speed jump ($\Delta U$) of $-50 \;\mathrm{km\;s^{-1}}$ isosurface (dark yellow) as a reference for the CME front, together with the HCS (gray). The heliographic rotating (HGR) coordinates are used. The background plane is at $X_\mathrm{HGR} = 3.5 \;R_\mathrm{s}$ and colored by the electron density scaled by heliocentric distance ($r^2N_\mathrm{e}$) on a logarithmic scale. 
    (d)(e) AWSoM-R modeled density cuts at $X_\mathrm{HGR} = 3.5 \;R_\mathrm{s}$ (cut 1) and $X_\mathrm{HGR} = 1.2 \;R_\mathrm{s}$ (cut 2), respectively, viewed from $50^\circ$ west of STA in longitude. These cuts are colored by the electron density scaled by heliocentric distance ($r^2N_\mathrm{e}$) on the same logarithmic scale, mainly showing the enhanced-density regions corresponding to the CME propagation fronts P1 and P2 in panel (b). The central red dashed circle marks the SC inner boundary at $r = 1.1 \;R_\mathrm{s}$ for reference.} \label{fig4:cmewl}
    \end{figure*}

\begin{figure*}[tp!]
    \centering{
    \includegraphics[width=1.0\textwidth]{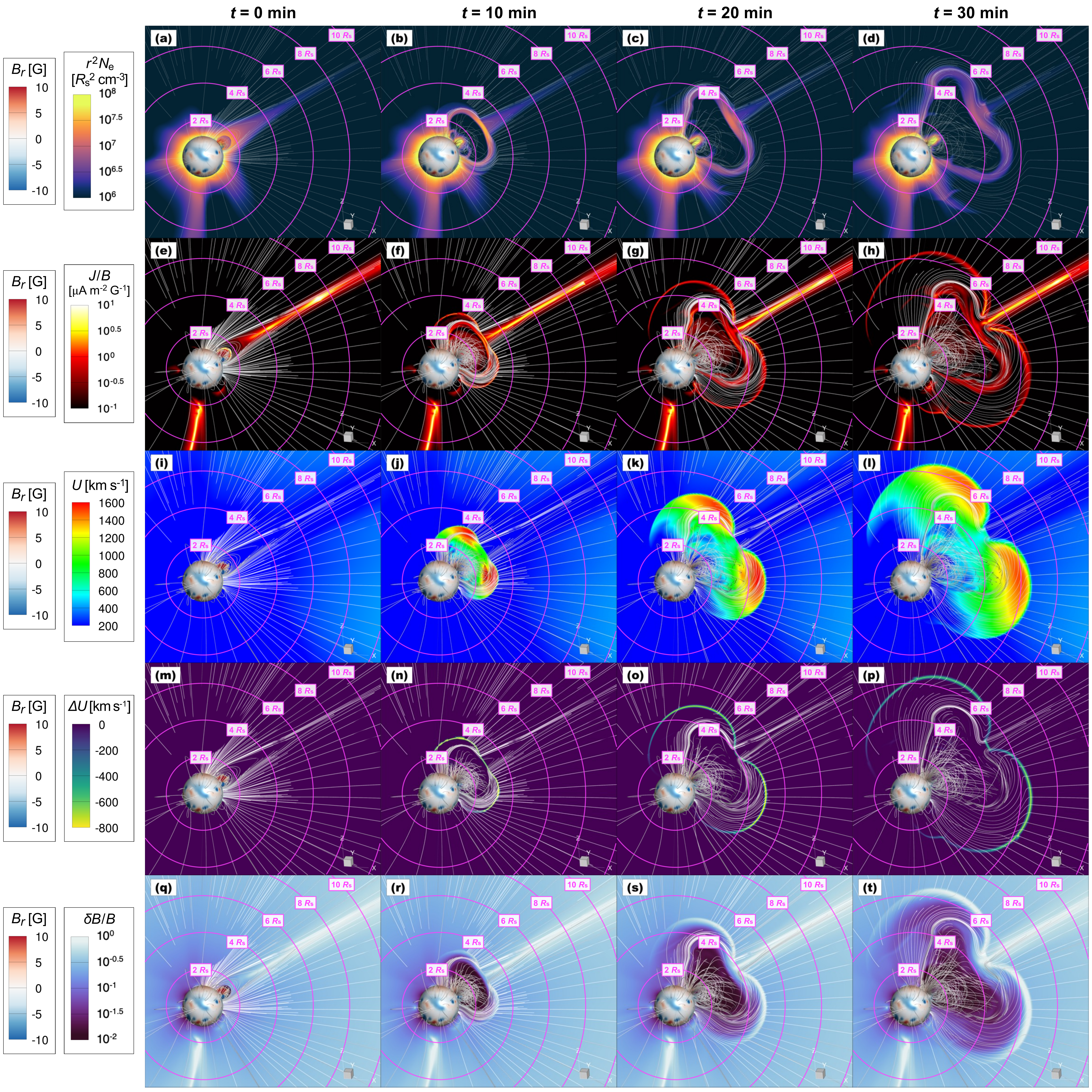}}
    \caption{3D evolution of the modeled CME during the first 30 minutes after the CME eruption. Columns show snapshots at $t = 0$, 10, 20, and 30 minutes. In each panel, the inner boundary of SC at $r = 1.1 \;R_\mathrm{s}$ and selected magnetic field lines are colored by the $B_r$ strength, and magenta concentric circles mark heliocentric distances every $2 \;R_\mathrm{s}$. The plane cut is taken through the source region and colored by different parameters in panels of different rows. 
    Panels (a)–(d) show the coronal density structure, represented by the electron density scaled by the heliocentric distance ($r^2 N_\mathrm{e}$). Panels (e)–(h) show the normalized current density ($J/B$), indicating sharp magnetic gradients and current-sheet structures. Panels (i)–(l), (m)–(p), and (q)–(t) show the CME-driven disturbance by the solar wind plasma speed ($U$), speed jump among adjacent cells ($\Delta U$), and the magnetic fluctuation ratio ($\delta B / B$) indicating the magnetic-field turbulence level, respectively.} \label{fig5:3d}
    \end{figure*}

    To model the CME eruption, we insert a force-imbalanced Gibson-Low flux rope configuration \citep{gibson1998time} with a spheromak-type magnetic field anchored at the inner boundary of the SC domain into the background solar wind. As listed in Table \ref{tab1:param}, the flux rope parameters are empirically chosen to reproduce the observed CME propagation direction, speed, and large-scale WL morphology \citep[e.g.,][]{borovikov2017eruptive, jin2017data}. The inserted flux rope then rises self-consistently and generates a CME in the modeled solar wind background (see Figures~\ref{fig4:cmewl} and \ref{fig5:3d}). Since Earth and STA have similar viewing perspectives during this event, here, we compare the simulated CME with STA/SECCHI/COR2 observations. 

    In Figure~\ref{fig4:cmewl}(a), the STA/COR2 observation exhibits two CME-associated features, labeled P1 and P2, divided by the streamer S2 and the neighboring structure S3. \textBoldmod{These two}{The corresponding P1 and P2} features are also recognizable in the synthetic WL image shown in Figure~\ref{fig4:cmewl}(b). \textBoldadd{For reference, the yellow dashed lines indicate the locations of S1–S3 identified in Figure~\ref{fig3:bgwl}, drawn along position angles of their maximum WL intensity in the pre-eruption corona. Because Figures~\ref{fig4:cmewl}(a) and \ref{fig4:cmewl}(b) display the ratio of the CME brightness to the pre-event background, the relatively stationary S1–S3 structures are not distinctly visible in these ratio images.} In addition, we notice a brightening in the STA/COR2 observation near the north pole, which is also captured in the synthetic image and is likely due to the interaction between the CME flank and streamer S1. Overall, the comparison between Figures~\ref{fig4:cmewl}(a) and \ref{fig4:cmewl}(b) shows that the synthetic WL image captures the direction, position, and large-scale morphology of the main CME-associated features. 
    To understand the underlying physical processes of these features, we further analyze the source regions of the WL signals associated with P1 and P2, as shown in Figures~\ref{fig4:cmewl}(c)–(e). Note that the viewing angle in these three panels differs from the STA/COR2 viewpoint\textBoldadd{: panels (c)–(e) are rendered in HGR coordinates from a direction $50^{\circ}$ west of STA in longitude, which better visualizes and separates the source regions associated with P1 and P2}. Figure~\ref{fig4:cmewl}(c) shows that the CME morphology is strongly affected by the HCS and splits into two components. We then identify the source regions of P1 and P2 by examining the enhanced-density structures along the corresponding lines of sight, and mark them on the two plane cuts shown in Figures~\ref{fig4:cmewl}(d) and \ref{fig4:cmewl}(e), respectively. 
    The spatial positions of these source regions suggest that P1 and P2 originate from the right and left components of the CME shown in Figure~\ref{fig4:cmewl}(c). This result indicates that features P1 and P2 in the synthetic WL image result from a splitting of the CME structure. Since similar features P1 and P2 are also present in the STA/COR2 observation, the comparison in Figures~\ref{fig4:cmewl}(a) and \ref{fig4:cmewl}(b) suggests that a similar CME splitting, likely driven by the interaction between the CME and the HCS/streamer structure, may have occurred in this event. The following analysis further examines this CME splitting in the simulation. 

    Figure~\ref{fig5:3d} shows the 3D evolution of the modeled CME during the first 30 minutes after the CME eruption onset. The first column shows that the initially inserted flux rope is located within the HCS, which is associated with an enhanced-density streamer structure, as indicated in panels (a) and (e) \textBoldadd{and consistent with the background solar-wind configuration and initial CME propagation direction shown in Figures~\ref{fig2:sw}(a) and \ref{fig2:sw}(b)}. During the early expansion of the flux rope, the lateral portions of the CME undergo expansion with relatively little constraint, while the central portion encounters the dense HCS/streamer structure and is partially impeded, as shown in the second and third columns of Figure~\ref{fig5:3d}. 
    By $t = 20$ minutes, this interaction has already deformed the CME front and led to the development of two distinct flanks. This asymmetric CME expansion further affects the formation and evolution of the CME-driven shock shown in panels (i)–(p), as well as the associated MHD turbulence level shown in panels (q)–(t), producing different shock strengths in different directions as discussed later in Section \ref{sec2.3:shk}. 
    We also note that CME and shock distortions by structured ambient solar wind have long been studied \citep[e.g.,][]{odstrvcil1996propagation, odstrcil1999distortion, prise2015analysis, wijsen2023effect}. Such CME–HCS interactions can further generate complex structures such as mesoscale flux ropes, enhanced current layers, and localized turbulent regions around the CME front \citep[e.g.,][]{geyer2023interaction, romeo2023near, manchester2025high}. 
    
    \subsection{Shock--Capturing and Magnetic Connectivity} \label{sec2.3:shk}
    
    \begin{figure*}[tp!]
    \centering{
    \includegraphics[width=0.925\textwidth]{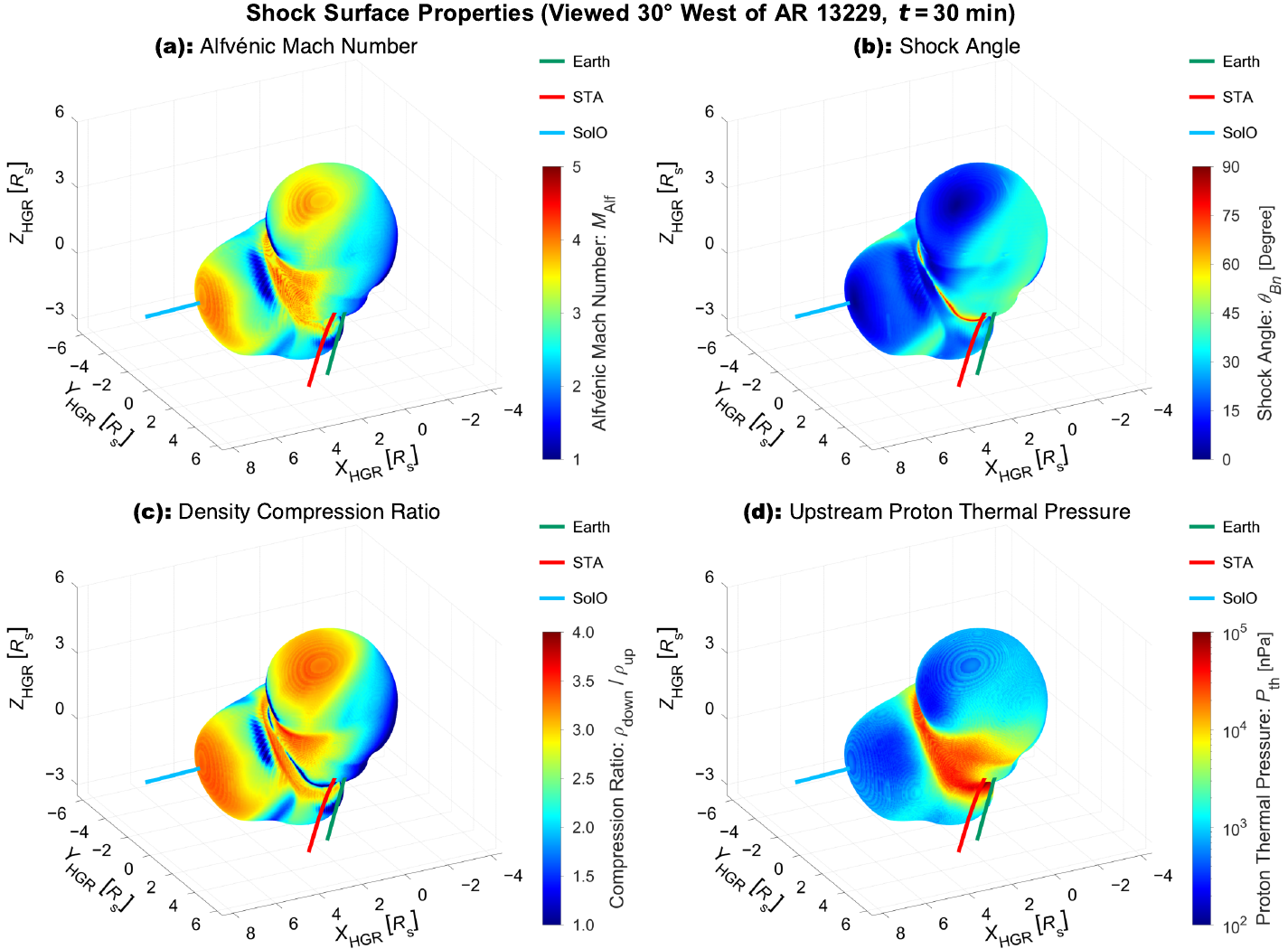}}
    \caption{CME-driven shock front at $t = 30$ minutes after the CME eruption, plotted in HGR coordinates and viewed from \textBoldmod{above}{30$^{\circ}$ west of} the CME source region (AR 13229) \textBoldadd{to more clearly show the observer-connected field lines and their intersections with the shock surface}. Panels are colored by (a) Alfv\'{e}nic Mach number ($M_\mathrm{Alf}$), (b) shock angle ($\theta_{Bn}$), (c) density compression ratio ($\rho_\mathrm{down}/\rho_\mathrm{up}$), and (d) proton thermal pressure upstream of the shock ($P_\mathrm{th}$). Colored curves in each panel are the magnetic field lines connected to Earth (green), STA (red), and SolO (blue)\textBoldadd{, and their intersections with the shock surface define the corresponding cobpoints}.} \label{fig6:shk}
    \end{figure*}
    
    To analyze the CME-driven shock and its magnetic connectivity to each observer, we apply a recently developed shock--capturing tool, which is embedded in AWSoM-R\textBoldrm{ and}\textBoldadd{,} described in Section 4.3 of \citet{liu2025physics}\textBoldadd{, and applied in recent studies \citep[e.g.,][]{chen2025evidence, liu2025modelingAGU, zhang2026radio}}. Starting from a longitude-latitude grid, the shock front is identified along each radial direction as the surface where the speed jump reaches its largest magnitude, and the local shock normal is determined from the shock surface geometry. For each point on the captured shock surface, we sample the upstream and downstream plasma states and derive a variety of shock parameters as shown in Figure~\ref{fig6:shk}. We then trace \textBoldadd{the} magnetic field lines \textBoldadd{in the AWSoM-R solution} from Earth, STA, and SolO back to the shock surface to determine their cobpoints, defined as the intersections between observer-connected field lines and the shock front. The concept of a cobpoint, short for ``Connecting with the OBserver Point", was first explicitly introduced in SEP modeling by \cite{heras1995three}. 
    
    \begin{figure}[tp!]
    \centering{
    \includegraphics[width=0.5\textwidth]{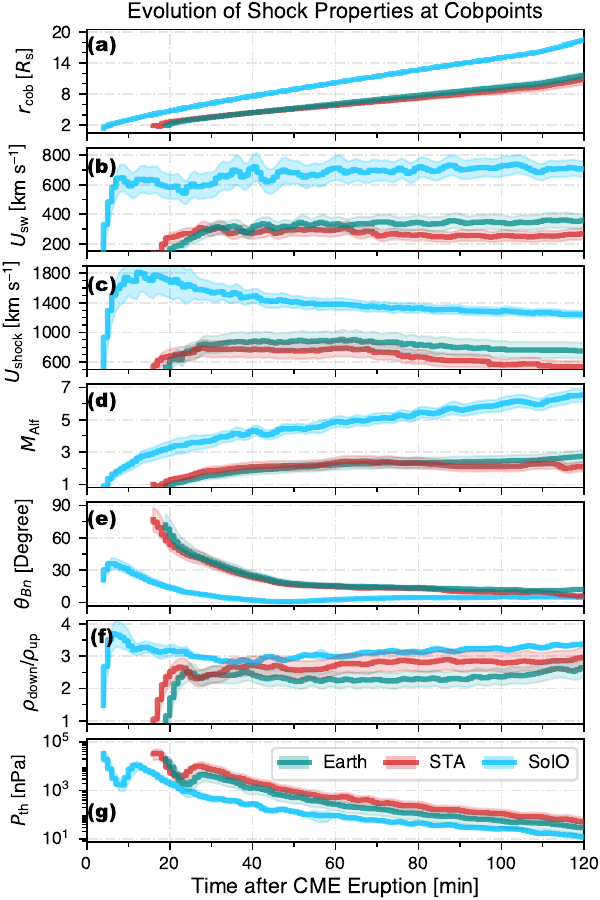}}
    \caption{Time evolution of the shock properties at the cobpoints of Earth (green), STA (red), and SolO (blue) during the first 2 hours after the CME eruption. Panels from top to bottom show (a) cobpoint heliocentric distance ($r_\mathrm{cob}$), (b) upstream solar wind speed ($U_\mathrm{sw}$), (c) shock wave speed ($U_\mathrm{shock}$), (d) Alfv\'{e}nic Mach number ($M_\mathrm{Alf}$), (e) shock angle ($\theta_{Bn}$), (f) density compression ratio ($\rho_\mathrm{down}/\rho_\mathrm{up}$), and (g) upstream proton thermal pressure ($P_\mathrm{th}$). 
    \textBoldadd{In each panel, solid curves show the values at the observer-connected cobpoints, and the light-colored shaded bands indicate the variability of shock properties sampled at the cobpoints of neighboring field lines within $10^{\circ}$ in both longitude and latitude around each observer.}} \label{fig7:cob}
    \end{figure}
    
    Figure~\ref{fig6:shk} shows the captured 3D shock surface colored by different properties. At $t = 30$ minutes after the CME eruption, the shock surface is already highly asymmetric: the two lateral portions of the erupting flux rope develop into two distinct shock flanks with separate local noses, while the central portion propagates more slowly between them, which corresponds to the CME–HCS interaction and flux rope deformation shown in Figures~\ref{fig4:cmewl} and \ref{fig5:3d}. 
    \textBoldadd{At this time, the simulation-derived cobpoints are located at $(3.38\;R_\mathrm{s},\, 71.0^{\circ},\, -4.5^{\circ})$ for Earth, $(3.48\;R_\mathrm{s},\, 64.0^{\circ},\, -3.5^{\circ})$ for STA, and $(5.95\;R_\mathrm{s},\, -3.5^{\circ},\, -2.0^{\circ})$ for SolO in HGR coordinates, given as the heliocentric distance, longitude, and latitude, respectively. Thus, the Earth and STA cobpoints remain close to each other, differing by only $0.10\;R_\mathrm{s}$ in radial distance, $7.0^{\circ}$ in longitude, and $1.0^{\circ}$ in latitude; the SolO cobpoint lies at a similar latitude but is located $\sim$2.5$\;R_\mathrm{s}$ farther from the Sun and $\sim$70$^{\circ}$ away in longitude.} 
    As shown in Figures~\ref{fig6:shk}(a)–(c), as well as in Figure~\ref{fig8:zcuts}(a) later, the magnetic field lines connected to Earth and STA intersect the shock on a weaker and more oblique flank, whereas the SolO-connected field line reaches a stronger shock region farther outward along the shock front, with a higher Alfv\'{e}nic Mach number, a smaller shock angle, and a larger compression ratio. This difference is further quantified by the time-evolving properties of the shock front at the \textBoldadd{simulation-derived} cobpoint \textBoldmod{of}{associated with} each heliospheric location shown in Figure~\ref{fig7:cob}. \textBoldadd{The light-colored shaded bands in Figure~\ref{fig7:cob} indicate the variability of shock properties sampled at the cobpoints of neighboring field lines within $10^{\circ}$ in both longitude and latitude around each observer. This $10^{\circ}$ window is chosen because it is comparable to the 1-day corotation-related displacement considered for the spacecraft locations in this event and consistent with the angular spacing used to initialize the magnetic field lines in M-FLAMPA (Section \ref{sec3.1:mflampa}). We also note that, although the large-scale HCS–SIR structure is constrained by the multi-spacecraft comparisons shown in Figure~\ref{fig2:sw}, uncertainties in the global MHD background solar wind \citep[e.g.,][]{jivani2023global, huang2024adjusting, huang2024solar} may affect the detailed magnetic connectivity and cobpoint properties. The shaded bands therefore provide a local estimate of this connectivity-related variability around each observer.} Compared with Earth and STA, \textBoldadd{regions near} SolO \textBoldmod{establishes}{systematically establish} an earlier magnetic connection to the shock by $\sim$15 minutes and generally samples a faster and stronger shock, characterized by a much larger shock speed, a higher Alfv\'{e}nic Mach number, and a larger density compression ratio during the early phase of this event. 
    
    For SEPs, this early phase is also when most suprathermal particles are injected. 
    Figures~\ref{fig6:shk}(d) and \ref{fig7:cob}(g) show the proton thermal pressure due to the background solar wind upstream of the shock. Spatial variations across the shock surface mainly reflect differences in the ambient solar wind, including the different heliocentric distances reached by different portions of the shock, and correspond to variations in the local injection efficiency in the SEP model, as described later in Section~\ref{sec3.1:mflampa}. Since the SolO shock cobpoint is associated with a higher shock speed and has moved farther away from the Sun than those of Earth and STA (Figure~\ref{fig7:cob}(a)), the upstream proton thermal pressure is correspondingly lower. 
        
    Later, as the CME flux rope propagates into the IH domain and evolves into an ICME, the associated ICME-driven shock is reported by DONKI\footnote{Details can be found in \url{https://kauai.ccmc.gsfc.nasa.gov/DONKI/view/IPS/23976/2} and \url{https://kauai.ccmc.gsfc.nasa.gov/DONKI/view/IPS/23980/1}. \label{ftn:arrival}} to arrive at Earth and STA at 18:43 UT and 14:30 UT on February 26, respectively, and is observed to arrive at SolO at $\sim$18:00 UT on February 26, as overplotted in Figure~\ref{fig2:sw}(e). In our simulation, the modeled interplanetary shock arrives at Earth, STA, and SolO at $\sim$15:00 UT, $\sim$10:30 UT, and $\sim$13:00 UT on February 26, respectively, showing agreement with the observed ICME shock arrivals within 5 hours, which is reasonable for global CME/ICME propagation simulations \citep[e.g.,][and references therein]{riley2018forecasting, chen2025decent, mays2025nasa}.

\section{SEP Simulations} \label{sec3:sep}
    
    Based on the MHD simulation and shock-connectivity analysis in Section \ref{sec2:mhd}, we further simulate the SEP distribution function from the SC to the IH domains using the Multiple-Field-Line Advection Model for Particle Acceleration (M-FLAMPA). We briefly describe the model in Section \ref{sec3.1:mflampa} and show the results with comparison to multi-spacecraft measurements in Section \ref{sec3.2:sepflux}. 
    
    \subsection{The M-FLAMPA} \label{sec3.1:mflampa}
    
    M-FLAMPA is a finite-volume SEP model that solves the Parker transport equation \citep{parker1965passage} for particle diffusive shock acceleration \citep{fermi1949origin, drury1983introduction} and transport along multiple magnetic field lines \citep{sokolov2004new, sokolov2024simultaneous, borovikov2018toward, zhao2024solar, liu2025physics, liu2026testbed}. 
    In this study, we adopt a numerical scheme based on the integral form of the Poisson bracket, which conserves the total number of particles throughout the M-FLAMPA simulation \citep{sokolov2023high, liu2025physics}. 
    In M-FLAMPA, we initialize 648 magnetic field lines uniformly on a sphere at $r = 2.5 \;R_\mathrm{s}$, covering $360^{\circ}$ in longitude and $\pm85^{\circ}$ in latitude, and trace them inward to the SC inner boundary at $1.1\; R_\mathrm{s}$ and outward to the IH outer boundary at $650\; R_\mathrm{s}$. These field lines are then advected through the coupled SC and IH domains using Lagrangian points, with a 2-minute coupling cadence between AWSoM-R and M-FLAMPA. 
    
    The injected seed particle density near the shock front is prescribed from the suprathermal tail of the solar wind \citep[e.g.,][]{gloeckler2003ubiquitous, fisk2006common} and is proportional to the local thermal pressure shown in Figures~\ref{fig6:shk}(d) and \ref{fig7:cob}(g). The suprathermal tail extends from the plasma thermal energy to the particle injection energy of 10 keV, and we apply a \textBoldadd{uniform} multiplier of 200 to the seed particle population in order to match the observed SEP intensities in multi-spacecraft comparisons. \textBoldadd{This global scaling adjusts the absolute intensity normalization but does not change the relative differences among the three spacecraft.} 
    
    Particle transport is treated along magnetic field lines, with the parallel diffusion coefficient representing pitch-angle-scattering transport along the mean magnetic field \citep[e.g.,][]{jokipii1966cosmic, zank2014transport}. In the far-upstream region of the shock, defined as beyond 1.05 times the shock distance, the diffusion coefficient is prescribed through a reference parallel mean free path of 0.5 au for protons with kinetic energy of $\sim$0.43 GeV\footnote{In this case, $pc=1$ GeV for protons, where $p$ and $c$ denote the proton momentum and the speed of light, respectively.} at $r = 1 \;\mathrm{au}$ \citep[see Equation~(27) of][]{li2003energetic} for this event. 
    Near and downstream of the shock, the diffusion coefficient is instead calculated from the local MHD turbulence \citep[see][and references therein]{borovikov2019toward}. The turbulence power spectral density is assumed to follow a Kolmogorov spectrum \citep{kolmogorov1941local}, with a correlation length of 0.005 au at $r = 1 \;\mathrm{au}$ \citep[e.g.,][]{kiyani2015dissipation, oughton2021solar} that scales linearly with the heliocentric distance toward the Sun, allowing enhanced particle trapping and scattering near the shock. For instance, as described below, Figure~\ref{fig8:zcuts}(b) shows the resulting MHD-turbulence-based parallel mean free path for the injected protons at 10 keV, with reduced values around the CME-driven shock corresponding to the strong turbulence shown in Figures~\ref{fig5:3d}(q)–(t). 
    In the present simulation, cross-field diffusion and drift effects are not included. Hence, the modeled SEP differences at different locations mainly reflect their different magnetic connections to the evolving CME-driven shock. 
    
    \subsection{SEP Fluxes and Multi-Spacecraft Comparison} \label{sec3.2:sepflux}
    
    \begin{figure*}[tp!]
    \centering{
    \includegraphics[width=1.0\textwidth]{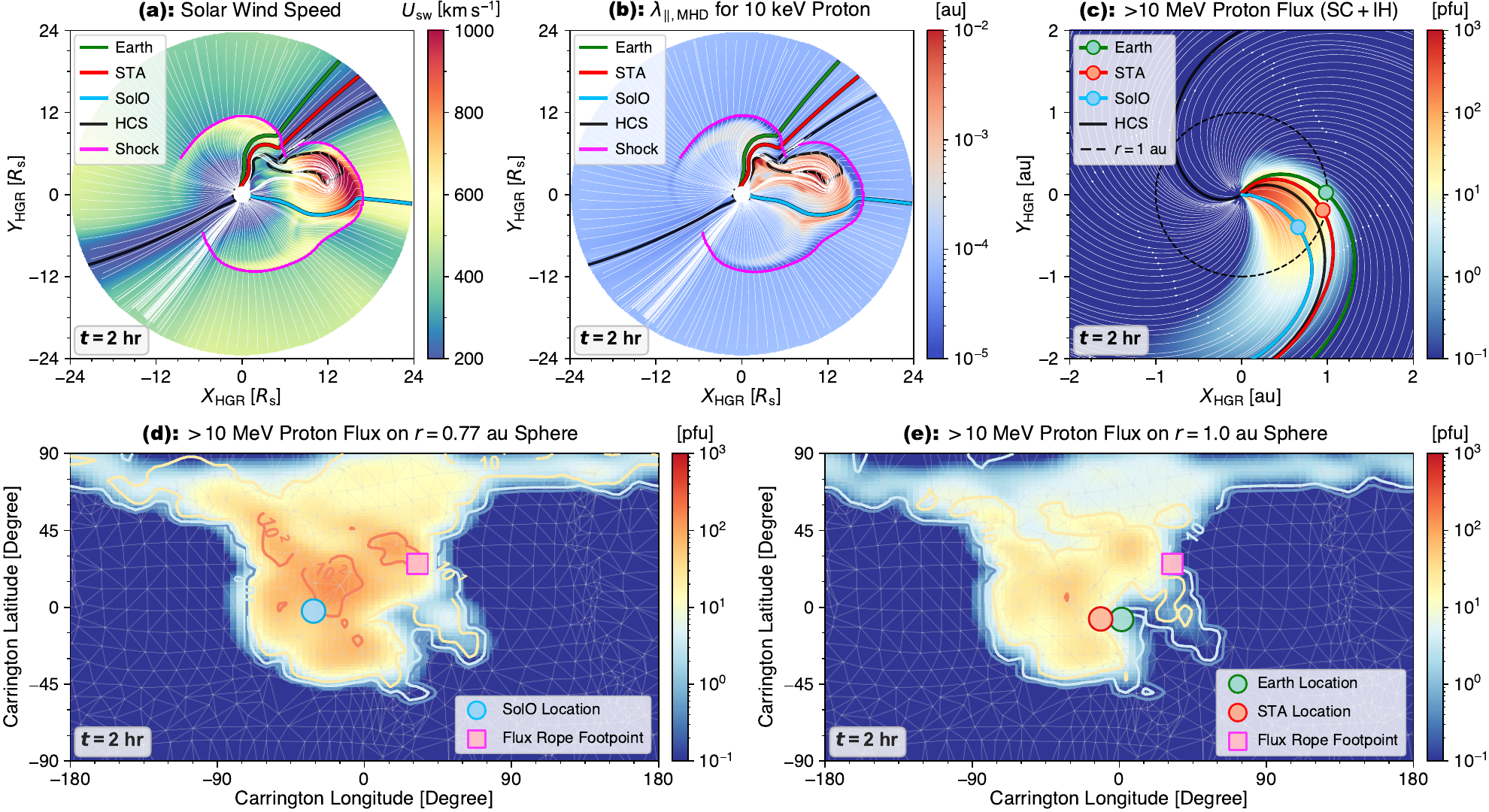}}
    \caption{Modeled solar wind properties and energetic proton flux distribution at $t = 2$ hours after the CME eruption. 
    (a) Solar wind speed ($U_\mathrm{sw}$) and (b) MHD-turbulence-based parallel mean free path ($\lambda_{\parallel, \,\mathrm{MHD}}$) for 10 keV injected protons in the solar equatorial plane of the SC domain. White curves indicate magnetic field lines, colored curves indicate field lines connected to Earth (green), STA (red), and SolO (blue), the black curve marks the HCS, and the magenta curve marks the captured shock front. 
    (c) Modeled $>$10 MeV proton flux in the combined SC and IH domains, with observer-connected field lines overplotted in the same style as in panels (a) and (b). The dashed black circle marks $r = 1$ au. 
    (d)(e) Modeled $>$10 MeV proton flux on spherical shells at $r = 0.77$ au and 1.0 au, respectively. The thin gray vertices correspond to the field line intersections with the spherical shell, and the edges illustrate the triangulation skeleton constructed by the Delaunay triangulation approach \citep{delaunay1934sphere} and used for interpolations. Circles mark the spacecraft locations using the same colors as in panels (a)–(c), and magenta squares represent the CME flux rope footpoint associated with the source region AR 13229 (see Figure~\ref{fig1:obs}(b)).} \label{fig8:zcuts}
    \end{figure*}
    
    In Figure~\ref{fig8:zcuts}, we plot the modeled solar wind properties, magnetic connectivity, and global SEP flux distributions at $t = 2$ hours after the CME eruption. 
    In the solar equatorial plane, the connectivity difference described in Section~\ref{sec2.3:shk} persists: Earth and STA field lines intersect a weaker flank of the CME-driven shock, whereas the SolO field line reaches a different and stronger shock region (see panel (a)). The SolO-connected region also has a smaller MHD-turbulence-based parallel mean free path near and downstream of the shock, indicating enhanced particle scattering and acceleration (see panel (b)). As a result, the modeled SEP flux is strongly structured by the shock geometry and magnetic connectivity. 
    
    Figure~\ref{fig8:zcuts}(c) shows the resulting $>$10 MeV proton flux in the solar equatorial plane from the SC to IH domains, while Figures~\ref{fig8:zcuts}(d) and \ref{fig8:zcuts}(e) show spherical cuts at $r = 0.77$ au and 1.0 au, respectively. 
    As expected from a nominal Parker-spiral magnetic field configuration, the high-flux regions are primarily located eastward of the eruption site (indicated by the magenta squares in Figures~\ref{fig8:zcuts}(d) and \ref{fig8:zcuts}(e); see also Figure~\ref{fig1:obs}(b)). 
    The spherical maps also show two enhanced-flux regions: one near the equator extending northwestward, corresponding to the orange region near and north of SolO in Figure~\ref{fig8:zcuts}(d), and another at higher northern latitudes close to the flux rope footpoint. These two regions are consistent with the deformed two-flank shock structure shown in Figures~\ref{fig5:3d}, \ref{fig6:shk}(a), and \ref{fig6:shk}(b). SolO lies close to the near-equatorial high-flux region (Figure~\ref{fig8:zcuts}(d)), whereas Earth and STA are located in regions with lower and intermediate energetic proton fluxes (Figure~\ref{fig8:zcuts}(e)). 
    
    \begin{figure*}[tp!]
    \centering{
    \includegraphics[width=1.0\textwidth]{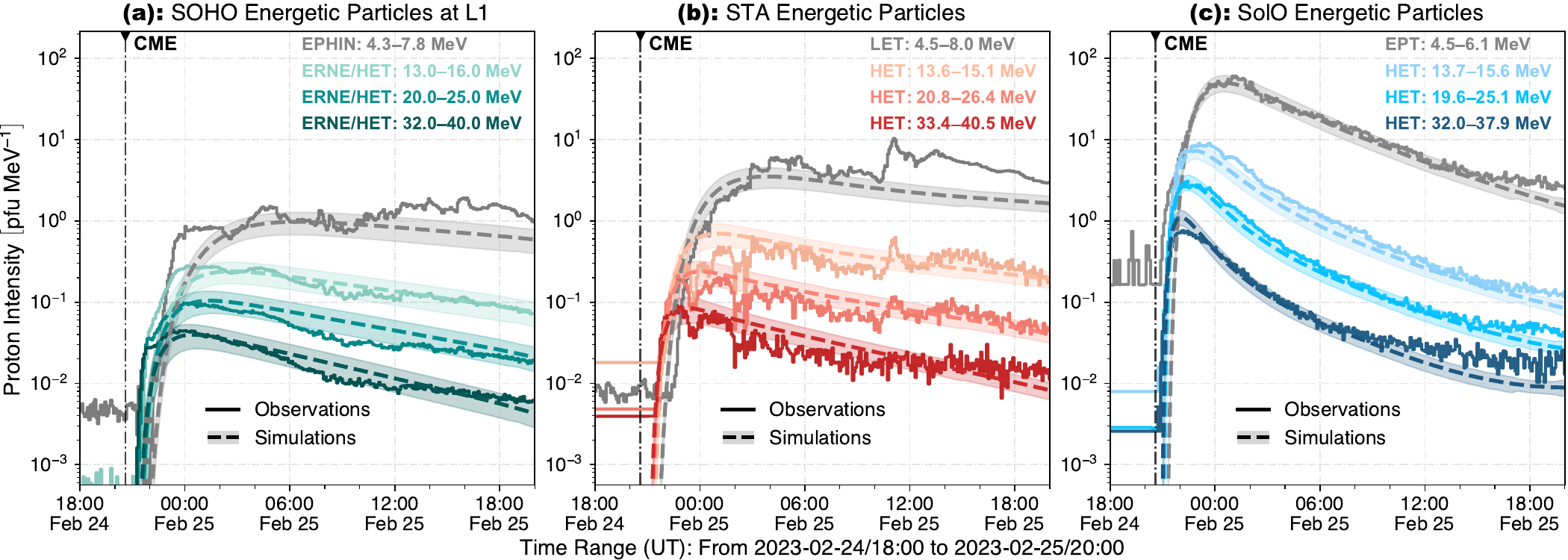}}
    \caption{Comparison between observed and simulated energetic proton time--intensity profiles at (a) L1/SOHO near Earth, (b) STA, and (c) SolO from 18:00 UT on 2023 February 24 to 20:00 UT on 2023 February 25. \textBoldrm{The solid and dashed curves indicate spacecraft observations and M-FLAMPA simulation results,} \textBoldmod{respectively}{Solid curves show spacecraft observations, while dashed curves show M-FLAMPA results for the field lines closest to the corresponding spacecraft}, in four representative energy channels listed in the upper-right legend, the same as those in Figures~\ref{fig1:obs}(c)–(e). \textBoldadd{Light-colored shaded bands, using the same colors as the corresponding energy channels, indicate the variation in the simulated intensities among neighboring modeled field lines within $10^{\circ}$ in longitude and latitude of each spacecraft.} The vertical dash-dotted black lines represent the CME eruption time.} \label{fig9:septime}
    \end{figure*}
    
    Figure~\ref{fig9:septime} shows the simulated energetic proton time--intensity profiles (dashed lines) and the observed intensities (solid lines) by, from left to right, SOHO, STA, and SolO for the first 24 hours of the event. 
    The simulated profiles are taken from the modeled field lines closest to each spacecraft, with longitudinal differences of $\lesssim$4$^{\circ}$ for L1/SOHO, $\lesssim$1$^{\circ}$ for STA, and $\lesssim$5$^{\circ}$ for SolO, and latitudinal differences within 2$^{\circ}$ for all three observers. 
    \textBoldadd{Considering the local flux variability associated with neighboring field lines, we estimate the uncertainty from the standard deviation of simulated intensities among modeled field lines within $10^{\circ}$ in both longitude and latitude around each observer. The light-colored shaded bands show the corresponding uncertainty ranges of the simulated intensities.} 
    Both simulations and observations exhibit clear energy-dependent dispersion from the onset to peak phases at all three spacecraft. Overall, the M-FLAMPA simulation reproduces the key ordering of SEP fluxes throughout this event: SolO observes energetic proton fluxes more than an order of magnitude higher than L1/SOHO and STA across the representative energy channels. \textBoldadd{The simulated variations among neighboring field lines are generally smaller than this inter-spacecraft intensity contrast. The uncertainty bands at SolO are also generally narrower than those at L1/SOHO and STA, likely because the SolO-connected field lines remain clustered around a stronger shock region, whereas the Earth- and STA-connected field lines intersect a more dynamic and nonuniform shock flank (see Figures \ref{fig6:shk} and \ref{fig7:cob}).} The SolO profiles in Figure~\ref{fig9:septime}(c) also show a more prompt onset-to-peak enhancement than those at L1/SOHO and STA, consistent with its earlier \textBoldadd{magnetic} connection to a stronger shock region shown in Figure~\ref{fig6:shk} \textBoldadd{\citep[see also][]{rouillard2016deriving, jarry2024evolution}}. 
    Some discrepancies can be found regarding the detailed onset and peak times, decay phases, and small-scale intensity variations. During the time interval shown in Figure~\ref{fig9:septime}, four additional small CMEs with speeds below $600 \;\mathrm{km \;s^{-1}}$ are reported by DONKI. These smaller eruptions may contribute to part of the observed fluctuations in particle intensities, especially in the lower-energy channels, which are not expected to be reproduced by the single CME simulation. 
    
    \begin{figure*}[tp!]
    \centering{
    \includegraphics[width=0.875\textwidth]{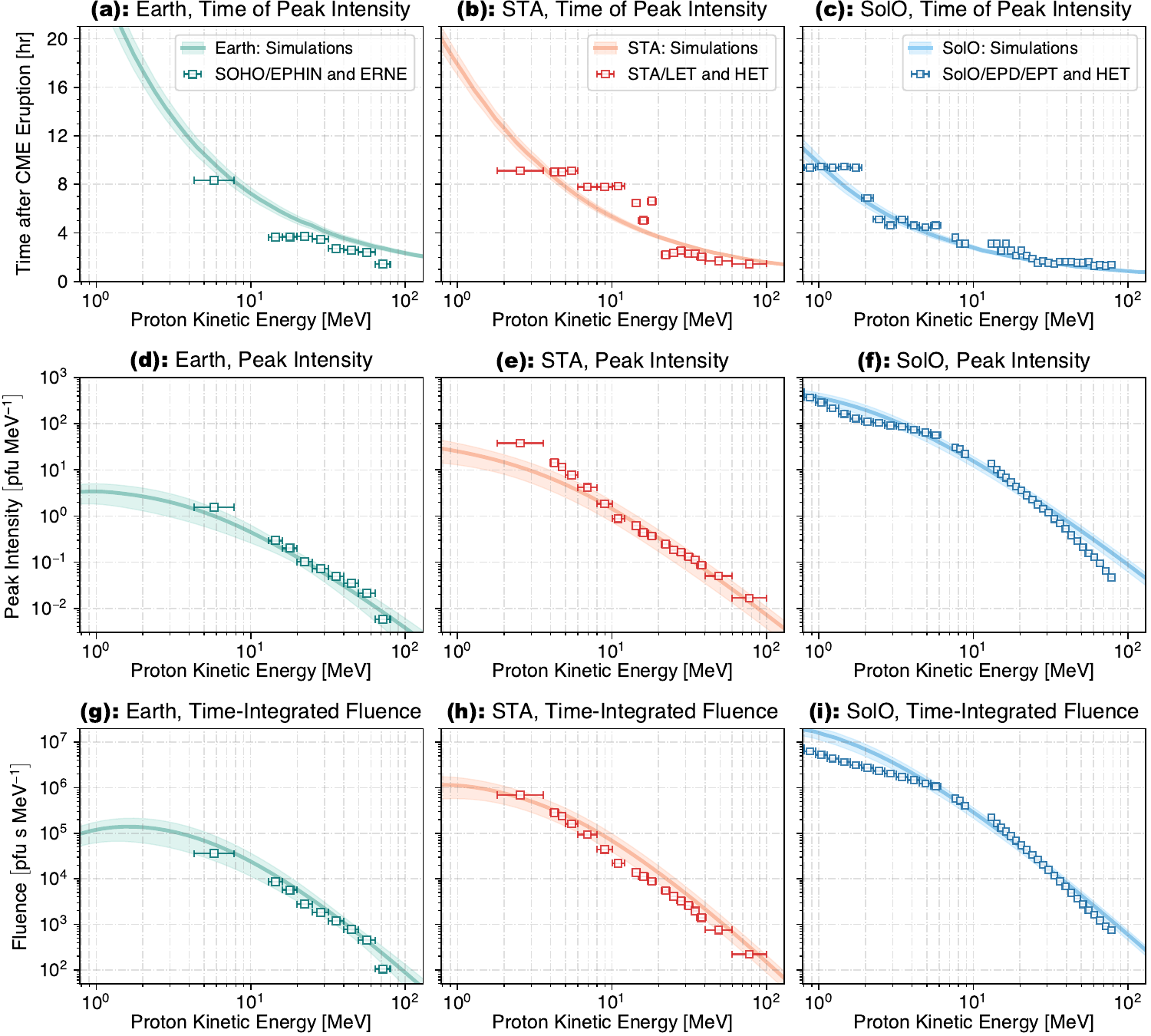}}
    \caption{Energy-dependent comparison between M-FLAMPA simulations and energetic-proton observations at Earth (green; left column), STA (red; middle column), and SolO (blue; right column). 
    Panels from top to bottom show (a)–(c) the time of peak intensity after the CME eruption, (d)–(f) peak intensity, and (g)–(i) time-integrated fluence as functions of the proton kinetic energy. 
    In each panel, the solid curve marks the simulation result \textBoldadd{for the field lines closest to the corresponding spacecraft, while light-colored shaded bands indicate the uncertainty associated with neighboring field lines within $10^{\circ}$ in both longitude and latitude, the same ones used and shown in Figure~\ref{fig9:septime}.} \textBoldmod{, and square}{Square} symbols mark spacecraft measurements where their energy bin widths are indicated by horizontal bars.} \label{fig10:mat}
    \end{figure*}
    
    In addition, Figure~\ref{fig10:mat} presents the model--observation comparison of peak times, peak intensities, and time-integrated fluences as functions of proton kinetic energy. In order to compute these fluences, we have considered the time interval from the CME eruption time to 19:36 UT on February 25, before the onset of another major CME eruption that significantly contributes to the later SEP intensities, as noted in Section~\ref{sec1.2:overview}. Besides, given that four smaller CME eruptions occurred within this event interval, we only consider peak times before 06:00 UT on February 25 for the observational data. The M-FLAMPA simulation results generally align with the observed energy-dependent trends at all three locations over the 1–100 MeV range, with peak-time differences generally less than $\sim$2 hours and peak-intensity and fluence differences within about half an order of magnitude. Specifically, M-FLAMPA captures the large relative enhancement in both peak intensity and fluence at SolO compared with Earth and STA. \textBoldadd{The shaded ranges in Figure~\ref{fig10:mat}, derived from the same neighboring field lines considered in Figure~\ref{fig9:septime}, further show that the modeled energy-dependent trends and larger particle fluxes at SolO remain robust against the local field-line variability.} Thus, the SEP simulations support the interpretation that the observed multi-spacecraft SEP flux differences are primarily attributed to different magnetic connections to the asymmetric CME-driven shock shaped by the HCS--SIR structure in the background solar wind, as discussed in Section \ref{sec2:mhd}. 
    

\section{Summary and Conclusions} \label{sec5:sumcon}

    In this study, we investigate the counterintuitive magnetic connectivity and SEP flux differences observed during the 2023 February 24 SEP event. In this event, Earth and STA near 1 au are nominally better connected to the CME eruption source region, while SolO at 0.77 au appears less favorably connected. Nevertheless, SolO observes SEP fluxes more than an order of magnitude higher than those at Earth and STA. This unusual behavior cannot be simply explained by nominal \textBoldrm{Parker-spiral} connectivity or radial distance effects alone, motivating a global analysis of the ambient solar wind, CME-driven shock, magnetic connectivity, and particle acceleration and transport. 

    Using SOFIE/AWSoM-R, we show that the ambient solar wind near Earth, STA, and SolO is strongly structured by the HCS and a nearby SIR. Although the three observers are close in heliolongitude, this SIR causes their observer-connected magnetic field lines to separate substantially in the SC domain. 
    The CME flux rope then propagates above the HCS and interacts with the surrounding HCS/streamer structure, leading to an asymmetric CME-driven shock with distinct flanks. Our shock--capturing analysis further shows that Earth and STA establish magnetic connections later and connect to a weaker and more oblique shock flank, while SolO connects earlier to a stronger shock region, with a larger Mach number, a higher density compression ratio, and thus more efficient particle acceleration. 

    We then simulate the SEP distribution function using SOFIE/M-FLAMPA. The modeled energetic proton flux distributions and time--intensity profiles reproduce the key observational feature during this event: SolO exhibits substantially higher energetic-proton fluxes than SOHO at L1 and STA across the representative energy channels. 
    The modeled energetic proton peak times, peak intensities, and event-integrated fluences are also consistent with all three spacecraft observations, further supporting the interpretation that the SEP flux differences observed by nearby multi-spacecraft are primarily caused by the different magnetic connections to the asymmetric CME-driven shock, shaped by the HCS--SIR structure in the background solar wind and CME flux rope deformation. 
    

\begin{acknowledgments}
    \textBoldadd{The authors sincerely thank the anonymous reviewer for the time and effort devoted to improving this paper.} 
    This work is supported by NASA Living With a Star (LWS) Strategic Capability project under NASA grant 80NSSC22K0892 (SCEPTER), NASA Space Weather Center of Excellence program under award 80NSSC23M0191 (CLEAR), and NSF ANSWERS grant GEO-2149771. 
    The authors express their sincere gratitude to the SOHO, STA, and SolO teams for making the observational data available. The authors acknowledge the use of the GONG synoptic map (footnote \ref{ftn:gong}), the LASCO (footnote \ref{ftn:lascocme}) and STA/SECCHI (footnote \ref{ftn:coracme}) CME lists, the DONKI database (footnotes \ref{ftn:donki} and \ref{ftn:arrival}), the ICME catalogs of Wind (footnote \ref{ftn:windcat}), STA (footnote \ref{ftn:stacat}), SolO (footnote \ref{ftn:solocat}), and ICMECAT (footnote \ref{ftn:icmecat}), as well as the ISWA system (footnote \ref{ftn:iswa}). The authors thank Dr. Spiro K. Antiochos from the University of Michigan for valuable discussions and insightful suggestions on this work. The authors also thank Dr. Patrick K\"uhl and Prof. Robert F. Wimmer-Schweingruber from the Christian-Albrechts-University Kiel in Germany for helpful guidance on the SolO/EPD/HET data. 
    Computational resources supporting this work are provided by the high-performance computing support from the Texas Advanced Computing Center (TACC) Frontera\footnote{\url{https://tacc.utexas.edu/}} at the University of Texas at Austin \citep{stanzione2020frontera}. 
\end{acknowledgments}



\bibliography{ref_paper}{}
\bibliographystyle{aasjournalv7}

\end{document}